\documentclass{clumps}
\usepackage{times}
\usepackage{graphicx}
\def\la{\hbox{{\lower -2.5pt\hbox{$<$}}\hskip -8pt\raise
-2.5pt\hbox{$\sim$}}}
\def\ga{\hbox{{\lower -2.5pt\hbox{$>$}}\hskip -8pt\raise
-2.5pt\hbox{$\sim$}}}

\begin{document}

\title{Detecting WIMPs in the Microwave Sky}
\author[1]{P. BLASI}
\affil[1]{INAF/Osservatorio Astrofisico di Arcetri, Largo Enrico
Fermi, 5 - 50125 Firenze, ITALY}
\author[2,3]{A. V. OLINTO}
\affil[2]{Department of Astronomy \& Astrophysics and Enrico Fermi
Institute, The University of Chicago, Chicago, IL 60637, USA}
\affil[3]{Center for Cosmological Physics, The University of Chicago,
Chicago, IL 60637, USA}
\author[2]{C. TYLER}

\pagestyle{plain}

\pagenumbering{arabic}

\titleheight{4cm} 
\maketitle

\begin{abstract}
The hierarchical clustering observed in cold dark matter
simulations results in highly clumped galactic halos. 
If the dark matter in our halo is made of weakly interacting
massive particles (WIMPs), their annihilation products should 
be detectable in the higher density and nearby clumps.  We
consider WIMPs to be neutralinos and calculate the synchrotron flux
from their annihilation products in the presence of the Galactic
magnetic field.  We derive a self-consistent emission spectrum
including pair annihilation, synchrotron self-absorption, and
synchrotron  self-Compton reactions. The resulting radiation spans
microwave frequencies that can be observed over the anisotropies in the
cosmic microwave background. These synchrotron sources should be
identifiable as WIMP clumps, either by their spatial structure
or by their distinctive radio spectrum.
\end{abstract}

\section{Introduction}

The density of dark matter in the Universe is observed via its
gravitational effects on galaxies and clusters of galaxies to
constitute about 30\% of the critical density of the
Universe. The nature of this dominant matter component is still 
unknown. Primordial nucleosynthesis and acoustic peaks in the
cosmic microwave background (CMB) constrain the
density of baryonic matter to be less than about 5\%
of the critical density, thus most of the dark matter is
non-baryonic. The leading candidate for the dark matter is the lightest
supersymmetric particle in supersymmetric extensions
of the standard model that can be stable by conservation of R-parity.
In most scenarios this weakly interacting massive particle (WIMP)
is the neutralino, $\chi$, which is a linear combination of the
photino, the zino, and the higgsinos (for a review, see
\citet{jkg96}).

Neutralinos may be detected directly as they traverse the Earth or
indirectly by the observation of their annihilation products. 
Direct neutralino searches are now underway in a number of low
temperature experiments with no consensus detection as of yet.
Indirect searches have been proposed both for gamma rays and
synchrotron emission from the annihilation of WIMPs in the Galactic 
center \citep{bgz92, bbm94}. The rate of neutralino annihilation is
proportional to the neutralino density squared ($\sim n_{\chi}^2$), 
therefore the strongest flux is expected to come from the
highest density regions such as the Galactic center. Depending on the
history of formation of the central black hole in our Galaxy,
the WIMP density may be strongly enhanced in the neighborhood of
the black hole. The density enhancement coupled with assumptions about
the local strength of the magnetic field can, in principle, constrain
viable models for neutralino masses and cross sections \citep{gs99,g00}.

Although promising, the proposal to focus indirect searches on the
Galactic nucleus is not free of uncertainties. The presence
of central cusps in large halos has been questioned by
observations \citet{sal00}. In addition, the large
fraction of baryonic matter as well as the history of the central
black hole are likely to alter the dynamics of the dark matter in
the Galactic center. Major mergers in the
central black hole's past will tend to decrease the dark matter
density. A clear picture of the magnetic structure around a
central black hole is also lacking.  In addition, when the
neutralino density becomes very high, the synchrotron signature is
strongly modified by synchrotron  self-Compton scatterings and
pair annihilation, as we discuss below and in \citep{AloGC}.
A cleaner dark matter annihilation signal may be
detected against a better understood background using the clumpy
nature of our halo.

Superimposed on the smooth component of CDM halos, high-resolution
simulations find a large degree of substructure formed by the constant
merging of smaller halos to form a present dark matter halo  (see, for
example, \citet{gmglqs}). The large number of clumps generated through
the hierarchical clustering of CDM comprises about $10 - 20\%$
of the total mass of a given halo. High density and nearby clumps in our
own halo  enhance appreciably the emission of gamma rays and
neutrinos from neutralino annihilation \citep{begu99,cm01}. As
these clumps reach closer to the Galactic plane, the Galactic
magnetic field strength increases drastically and the synchrotron
signal from the charged products of neutralino annihilation
intensifies considerably.

Here, we calculate in detail the synchrotron radiation of
electrons and positrons generated as decay products of WIMP
annihilation in the Galactic magnetic field. We show that the
synchrotron emission can provide a crucial test of WIMP models, since
the predicted fluxes are in the microwave region and exceed the signal
of CMB anisotropies at some frequencies. The detection of this excess
radiation from small angular size regions in the sky may provide the
first signal of WIMP dark matter.

In section 2, we explain the calculation procedure.  In sections
3 and 4 we present results for two different clump density profiles.
Section 5 is the conclusion, with prospects for detection.

\section{Synchrotron from Dark Matter Clumps}

\subsection{Dark Matter Clump Structure}

The density profile of dark matter halos has yet to be precisely
determined. The possible profiles for the smooth dark matter component
on the scales of galaxies is reasonably well constrained by observations
of galactic dynamics together with CDM simulations. In contrast, the
structure of the smaller dark matter halos (i.e., clumps) is only
constrained by theoretical arguments and numerical simulations. Our
Galactic halo can be described by both a smooth large scale
component plus a distribution of clumps which are the smaller
halos that have fallen into the Galactic potential well as it
formed from lower mass objects. 

In order to derive the synchrotron emission from neutralino
annihilation in the dark matter clumps of our Galaxy, we need to model
both the smooth component of the halo as well as the individual density
profile of  the clumps.  We consider the halo of the Galaxy to be well
described by the Navarro, Frenk, and White (NFW) density
profile~\citep{nfw} which we describe below. We model the individual
clumps of dark matter in our halo by two choices of CDM halo
profiles that span the range of reasonable slopes: the NFW profile
and the singular isothermal sphere (SIS). These two models bracket
a range of possible clump density profiles.

The NFW profile arises from CDM simulations and is 
consistent with observations of a number of galaxies. The dark matter
density as a function of radius from the center of the particular halo
is given by:
\begin{equation}
\rho(r) = m_{\chi} n_0 \left({r \over r_c}\right)^{-1} \left[{1 +
{r\over r_c}}\right]^{-2}  \ ,
\label{NFW_profile}
\end{equation}
where $r_c$ is a core radius and $m_{\chi}$ is the neutralino mass.
This profile gives a broad core region with a gently sloping
profile inside ($\rho \sim r^{-1}$) and a steeper slope outside
$r_c$ ($\rho \sim r^{-3}$). In general,  $n_0$ and  $r_c$ are fixed by
the total mass of the halo inside some radius $R_{\rm halo} \gg
r_c$ and the density at some particular point (or equivalently the
velocity dispersion at that point). To set these parameters for
the smooth component of our Galactic halo, we chose
$r_c = R_{\rm halo}/10$ where 
$R_{\rm halo}$ is the radius within which lies most of our halo's
mass, 
$M_{\rm halo}(r\le R_{\rm halo}) = 2 \times 10^{12} M_{\odot}$. In
addition, we set the  density in the solar neighborhood from the
local velocity dispersion to be
$\rho_{\odot}
\equiv \rho (8.5 \, {\rm kpc}) = 6.5 \times 10^{-25} {\rm g/cm}^3$ 
which fixes all  parameters.

For each individual clump described by the NFW profile, we fix 
$r_c$ and $n_0$ by setting $r_c = R_{c}/10$ where $R_{c}$ is the
radius at which the density of the clump equals that of the smooth
host halo density. The total
mass of the clump is then determined by the mass inside $R_{c}$.

\begin{figure}[t]
\vspace*{2.0mm}
\includegraphics[angle=0,width=8.3cm]{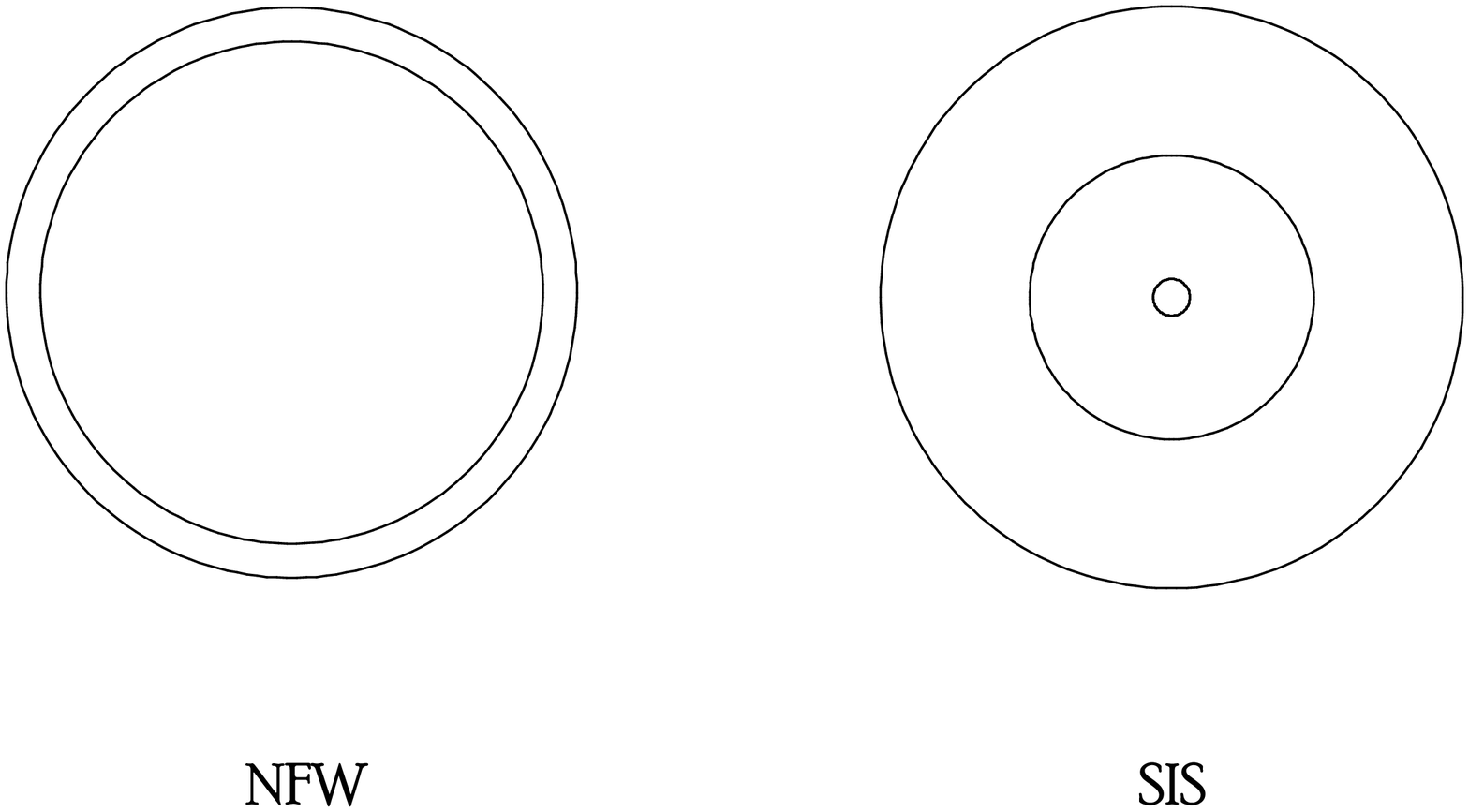}
\caption{Anatomy of a WIMP clump.  Radial distances are drawn
on a logarithmic scale, with the center chosen to be 
$10^{13}$~cm.  The NFW clump (left) shows $r_c$
inside $R_{\rm cl}$.  The SIS clump (right) shows, from
inner to outer, $R_{\rm min}$, $R_{1/2}$,
and $R_{\rm cl}$.  Both clumps
are $10^8~M_{\odot}$, with $R_{\rm cl} \sim 1$~kpc.}
\label{fig_anat}
\end{figure}

In order to study the range of possible clump radial profiles, we also
consider the case of clumps described by singular isothermal spheres
(SIS). The density profile in this case is given by:
\begin{equation}
\rho(r) = m_{\chi} n_0 \left({r \over r_c}\right)^{-2} .
\label{SIS_profile}
\end{equation}
By contrast with NFW, the SIS profile provides a steeper slope,
leading to a sharp peak density at the center of the clump.
The SIS model describes  a collection of self-gravitating
collisionless particles. If dark matter
clumps lack baryons, then this profile is likely to be more
appropriate.  The parameters $r_c$ and
$n_0$ are fixed by the total clump mass and the halo density at
the edge.  Figure~\ref{fig_anat} sketches both NFW and SIS
clumps ($R_{\rm min}$ and $R_{1/2}$ are described in section 4).

In addition to the specific halo and clump profiles, we need
a model for the distribution of clumps in the halo.  Following
\citet{bs00}, clumps of mass $m$ and position $r$ are
distributed according to
\begin{equation}
n_{\rm cl}(r,m) = n_{\rm cl,0}
\left(\frac{m}{M_H}\right)^{-\alpha}
\left[1+\left(\frac{r}{r_c^{\rm cl}}\right)^2\right]^{-3/2},
\label{clumpdist}
\end{equation}
where $n_{\rm cl,0}$ is a normalization constant and $r_c^{\rm cl}$
is the core of the clump distribution. A value of $\alpha\sim 1.9$
fits well the simulations in \cite{gmglqs}, in which
a halo with $M_H\approx 2\times 10^{12}~M_\odot$
contains about $500$ clumps with mass larger than $M_{\rm cl,min} \sim
10^8~M_{\odot}$.  The presence of such dark substructure gains
some support from recent analyses of strong gravitational lens
systems (see, for example,~\citet{mc01}), where dwarf galaxy
satellites and globular clusters are insufficient to affect the
lensing, but dark subhalo structures do alter the flux ratios
in multiply imaged systems.

The halo representations generated here
extend the clump distribution to a lower mass cutoff of
$10^7~M_{\odot}$. In principle, lower mass clumps will be present in
simulations with larger dynamical range. For the case of gamma-ray
emission, \cite{cm01} find that little additional flux 
arises if the minimum mass cutoff is lowered. A more detailed
study of lower mass clumps will be described in \cite{abot02}.

\subsection{Generating Synchrotron Radiation}

Neutralinos in DM clumps annihilate mostly into quark-anti-quark
pairs which hadronize mostly into pions. The neutral pions give
rise to the gamma-ray emission via $\pi^0 \rightarrow \gamma
\gamma$; while charged pions decay into charged leptons: 
\begin{equation}
\pi^+ \rightarrow \mu^+ \nu_{\mu} \,\, ,
\pi^- \rightarrow \mu^- \bar{\nu}_{\mu} \,,
\label{pidecay}
\end{equation}
and muons subsequently decay into electrons and positrons: 
\begin{equation}
\mu^+ \rightarrow e^+ \bar{\nu}_{\mu} \nu_e \,\, ,
\mu^- \rightarrow e^- \nu_{\mu} \bar{\nu}_e \,.
\label{mudecay}
\end{equation}
Neutrinos from this cascade process may eventually be
detectable, while electrons and positrons can be observed via
synchrotron radiation in the presence of magnetic fields.

The number of electrons and positrons produced at each energy
resulting from a single $\chi \bar{\chi}$ annihilation, ${dN_e \over
dE_e}$,  can be calculated via:
\begin{equation}
{dN_e \over dE_e} = \int_{E_e}^{m_{\chi} c^2}
\int_{E_{\mu}}^{E_{\mu}/\bar{r}} W_{\pi}
{dN_{\mu}^{(\pi)} \over dE_{\mu}}
{dN_e^{(\mu)} \over dE_e}\,dE_{\pi}\,dE_{\mu}\,,
\label{fragtot}
\end{equation}
where $\bar{r} \equiv (m_{\mu} / m_{\pi})^2$ and
\begin{equation}
W_{\pi} = {15 \over 16}
\left( {E_{\pi} \over m_{\chi} c^2} \right)^{-3/2}
\left( 1 - {E_{\pi} \over m_{\chi} c^2} \right)^2\,.
\label{Wpi}
\end{equation}
This form for the pion spectrum is the result of the fragmentation 
and  hadronization of partons, as found in \cite{hill}.
More sophisticated fits can be used but for the present work
eq. ({\ref{Wpi}) is sufficient. It is worth noting the strongly
non-thermal form of the fragmentation spectrum, which is 
instrumental in distinguishing DM annihilation from  the
background provided by the CMB radiation. In the integral
(\ref{fragtot}), the pion decay generates the following muon
number per unit energy:
\begin{equation}
{dN_{\mu}^{(\pi)} \over dE_{\mu}} =
{1 \over E_{\pi}}
{m_{\pi}^2 \over m_{\pi}^2 - m_{\mu}^2}
\label{fragpi}
\end{equation}
and the muon decay gives
\begin{equation}
{dN_e^{(\mu)} \over dE_e} = {2 \over E_{\mu}}
\left[ {5 \over 6} - {3 \over 2} \left( {E_e \over E_{\mu}}
\right)^2 + {2 \over 3} \left( {E_e \over E_{\mu}} \right)^3
\right]\,.
\label{fragmu}
\end{equation}
At low energies, eq.~(\ref{fragtot}) behaves as
\begin{equation}
{dN_e \over dE_e} \sim E_e^{-{3 \over 2}} ,
\label{dNdEsimp}
\end{equation}
and goes to zero as  $E_e$ reaches the maximum energy which
is of the order of $m_{\chi}$.

The total injection rate of electrons and positrons by neutralino
annihilations is then given by
\begin{equation}
q_e (E_e) = n_{\chi}^2 \langle \sigma v \rangle_{\chi {\bar \chi}}
\left(\frac{dN_e}{dE_e}\right) \,;
\label{qe}
\end{equation}
where $\langle \sigma v \rangle_{\chi
\bar{\chi}}$ is the annihilation rate.  Depending
on the specific neutralino model, the annihilation rate
varies around a fiducial value of $\langle \sigma v
\rangle_{\chi \bar{\chi}} \simeq 3 \times 10^{-29}~{\rm cm^3/s}$,
with a wide spread. The electron-positron population at any point
in space can be written as
\begin{equation}
\frac{dn_e}{dE_e} \simeq q_e \tau,
\label{dndE}
\end{equation}
where $\tau$ is the average lifetime of the electron or positron.  
For the cases considered below, the electron or positron gyroradius  is
usually much smaller than the size of the emitting region. It is a good
assumption to calculate the emission of electrons and positrons as
coming from their original position.
$\tau$ is the timescale over which electrons and positrons radiate
away most of their energy.

Depending on  the density profile of the clump, the
electron or positron energy, $E_e$, and the position within a
clump,  $\tau$ is determined by the shortest among the following
timescales: the energy loss timescale for (1) synchrotron
radiation, (2) inverse Compton scattering (ICS) off the cosmic
microwave background (CMB), (3) ICS off the local synchrotron
photons, and (4) the possibility of electron-positron
annihilation. In the NFW case discussed below, the synchrotron
emission is the main energy loss process. The SIS case has
different regimes with the central regions being dominated by ICS
off the generated synchrotron photons. We discuss each case in
more detail below.

The total synchrotron power \citep{rl},
\begin{equation}
{dE_{{\rm syn}} \over dt} = 4 \times 10^{-21} B_{\mu}^2 E_e^2
\,\,\,{\rm erg/s}\,,
\label{dEdtsyn}
\end{equation}
where $B_{\mu}$ is the magnetic field strength in $\mu$Gauss and
$E_e$ is the electron energy in GeV. In our Galaxy, estimates of
$B_{\mu}$ vary between 3 and 6. If we choose $E_e \sim  m_{\chi}
\sim 100$ GeV, the typical synchrotron power becomes $4\times
10^{-17} \rm{erg/s}$. The  peak frequency,
\begin{equation}
 \nu_{{\rm max}} = 3.7 \times 10^{6} B_{\mu} E_e^2
\,\,\,{\rm Hz}\,,
\label{numax}
\end{equation}
in the Galactic $\mu$Gauss fileds, indicate that the synchrotron
emission should be observable below about 50-100 GHz.

In the numerical calculations of individual clump signatures discussed
below, we model the Galactic magnetic  field $B(r,z)$ using 
\citet{st97}, where here $r$ and $z$ are Galaxy-centered cylindrical
coordinates. We also assume for simplicity that the synchrotron
radiation  is mainly produced at the peak frequency defined above.

If the only significant loss process is the emission of synchrotron
radiation, then
\begin{equation}
\tau \simeq {E_e \over dE_{\rm syn}/dt}\,.
\label{lifetime}
\end{equation}
The synchrotron emissivity from
annihilating neutralinos is then
\begin{equation}
j_{\nu} \simeq \frac{dn_e}{dE_e} \frac{dE_e}{d\nu}
\frac{dE_{{\rm syn}}}{dt}
\,\,\,\frac{{\rm erg}}{{\rm cm^3\,s\,Hz}}\,.
\label{jnu}
\end{equation}

In some regions of DM clumps, other processes affect the
electron-positron number density per energy  more strongly than the
synchrotrom emission. In the SIS case, the most significant process is
inverse Compton scattering (ICS) off the generated synchrotron photons.
Other effects come from the $e^+ - e^-$ pair annihilation and
synchrotron self absorption. We discuss these as they become
relevant in section 4.

\subsection{Comparing with the CMB}

As discussed above, the range of frequencies where synchrotron emission
from DM clumps may be observed peaks around $\sim$ 40 GHz
$(m_\chi/100 {\rm GeV})^2$. This frequency range is the focus of
intensive studies of the structure of  cosmic microwave background
(CMB) radiation. The CMB signals are well understood and well
measured and give the opportunity to search for the DM clump
emission. We focus our attention on the band between $10-400$~GHz,
which should be relatively free of contaminating CMB foregrounds
\citep{teho00}, and which is where the most sensitive CMB
experiments are planned to operate.

Only highly peaked SIS clumps may reach the levels of the CMB
emission. In general, the annihilation signal is comparable in
flux to the CMB anisotropies, which are at the level of  $\sim 10^{-5}$
of the CMB. In what follows, we assume that the CMB anisotropies
have a blackbody spectrum.

The best spectral data on the CMB come from the COBE FIRAS
instrument~\citep{fix96}.  The reported uncertainties are of  
order of $20~{\rm kJy~sr^{-1}}$, above 68 GHz  and only
above $7 {\rm ^o}$ angular scales.  The COBE DMR
instrument~\citep{smoot92} used 31, 53, and 90 GHz, and
found consistent anisotropies (of higher order than dipole)
at the $1.1 \times 10^{-5}$ level, but again on $7 {\rm ^o}$
scales.  A combined FIRAS and DMR analysis~\citep{fix97}
sets the anisotropy spectrum at about $5 \times 10^{-5}$
of the cosmic monopole spectrum for these large scales.

Our interest is in pixel sizes typical of upcoming
CMB probes.  The MAP mission~\citep{map00}
is typically 18~arcmin and the Planck
Surveyor~\citep{planck} is 5~arcmin, so we choose
a nominal pixel size of $10 \times 10$~arcmin
(the smaller the pixel, the brighter the clump, because
one can aim at the central cusp where the neutralino
annihilations are most rapid).  The BOOMERANG~\citep{boom00}
and MAXIMA-1~\citep{max00} experiments employed this resolution,
and they report $2.6 \times 10^{-5}$ fluctuations at $1^{\rm o}$
(BOOMERANG) and $1.7 \times 10^{-5}$ at 14.4~arcmin (MAXIMA-1),
with both quoted at 150~GHz.

On the lower frequency side, some of
the best measurements come from the VLA, probing very small
angular scales (where primordial plasma fluctuations become
damped during the nonzero timescale of recombination).  They
report $2 \times 10^{-5}$ at $8.4$~GHz~\citep{vla97,vla93}.

The VLA measurements compare meaningfully with clumps
that appear as point sources.  For extended sources, is
seems best to blend the various experimental results, keeping
in mind the pixel size of interest.  Fortunately, for all
scales, the data are consistent to within an order of magnitude.
We choose $\Delta I_{\nu} /  I_{\nu} \simeq 2 \times 10^{-5}$
as the limit for detection of deviations distinct
from those inherent in the CMB, with the intended caveat
that actual detections made in this way deserve further
scrutiny regarding the precision of the CMB blackbody.

Then to summarize, our criteria for clump detection over the
CMB are as follows:  Between 10 and 400~GHz, the center
pixel of the clump (taken to be $10 \times 10$~arcmin)
must exceed $2 \times 10^{-5}$ times $I_{\nu,{\rm CMB}}$
in the same pixel solid angle.

\section{NFW Clumps}

The NFW density profile, given in eq.~\ref{NFW_profile},
has a lower density central cusp when compared with a
SIS clump. This makes the NFW case simpler:  absorption of
synchrotron photons by relativistic electrons is unimportant and
$e^{\pm}$ pair annihilation occurs on a longer time scale
than synchrotron radiation, so it can be ignored as well.
We also neglect the presence of any constant density core,
as this region is in general too small to be detected.

In order to derive the NFW clump signature, it is sufficient to evaluate
the integral for the flux observed on Earth, from a
clump at distance $d$:
\begin{equation}
I_{\nu} = {1 \over 4 \pi d^2} \int_0^{R_{\rm cl}} 4 \pi r^2
j_{\nu}\,dr\,\,\,\frac{{\rm erg}}{{\rm cm^2\,s\,Hz}}\,,
\label{Inu_NFW}
\end{equation}
using eq.~\ref{jnu}.  Because $r_c$ is relatively large
($\sim$75~pc for a $10^8~M_{\odot}$ clump), NFW clumps will often
have observable angular sizes.  

The detection of a NFW clump can be made by observing both the
spectrum of the synchrotron emission as well as the detail
angular profile which is resolvable for NFW clumps. In 
figure~\ref{fig_specNFW}, we plot the spectrum of a specific NFW clump
with mass $10^8~M_{\odot}$ located at Galactocentric coordinates
[-4,0,0]~kpc (the Sun is at [-8.5,0,0]), and with the following base
properties:  $\langle \sigma v
\rangle_{\chi \bar{\chi}} = 3 \times 10^{-29}~{\rm cm^3/s}$ and
$m_{\chi} = 100$~GeV.  The thick line gives this clump's
spectrum from its central pixel, and the thin solid line
shows the CMB anisotropies (taken as $2 \times 10^{-5} 
I_{\nu,{\rm CMB}}$) in the same pixel solid angle. The other lines show
the effect of modifying the mass of the clump by a factor of 10 (dashed
line), increasing the cross section by a factor of 100  (dotted
line), and changing the neutralino mass by a factor of 10
(dot-dashed line) or a factor of 100 (dot-dot-dashed line).  One
can see that raising $m_{\chi}$ results in a higher cutoff
frequency, so that if a bright enough annihilating neutralino
clump were observed, its spectral cutoff would immediately reveal
$m_{\chi}$.  On the other hand, changes to $M_{\rm cl}$ and
$\langle \sigma v \rangle_{\chi \bar{\chi}}$ are degenerate, so
some other means is necessary to distinguish between the two, such
as gravitational lensing, or dynamical effects on nearby stars.

\begin{figure}[t]
\vspace*{2.0mm}
\includegraphics[angle=-90,width=8.3cm]{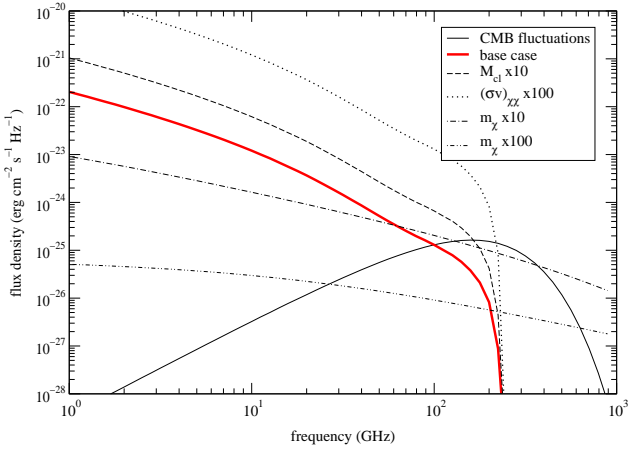}
\caption{Spectrum of synchrotron emission from an example NFW clump,
compared with the CMB anisotropies (thin line) in the same solid
angle as the clump.  This clump is chosen to lie
at Galactocentric coordinates [-4,0,0] kpc, 4.5 kpc away from us.
The heavy solid line is the
base case, where $M_{\rm cl} = 10^8~M_{\odot}$,
$\langle \sigma v \rangle_{\chi \bar{\chi}} = 3 \times
10^{-29}~{\rm cm^3/s}$, and $m_{\chi} = 100~{\rm GeV}$.  The dashed
line increases $M_{\rm cl}$ by 10 times; the dotted line increases
$\langle \sigma v \rangle_{\chi \bar{\chi}}$ by 100 times;
the dot-dashed line increases $m_{\chi}$ by 10 times; the
dot-dot-dashed line increases $m_{\chi}$ by 100 times.}
\label{fig_specNFW}
\end{figure}

\begin{table}
\begin{center}
\begin{tabular}{|l|rrrrrr|} \hline
{\bf NFW} & {\bf 1} & {\bf 2} & {\bf 3} & {\bf 4} 
& {\bf 5} & {\bf 6} \\ \hline
observable & 282 & 264 & 289 & 261 & 296 & 265 \\
outshine CMB & 0 & 0 & 0 & 0 & 0 & 0 \\
$> 30^{\rm o}$~latitude & 64 & 71 & 59 & 54 & 60 & 68 \\
within 3~kpc & 9 & 8 & 2 & 5 & 5 & 7 \\
\% of sky & 0.09 & 0.06 & 0.06 & 0.05 & 0.07 & 0.06 \\
\hline
\end{tabular}
\end{center}
\caption{Some results for NFW clumps in six halo realizations, each
set with $M_{\rm cl,min} = 10^7~M_{\odot}$, yielding 3972 clumps.
The following information on {\it observable} clumps is given:
number of them, number which outshine the CMB (rather than just
the CMB anisotropies), number above $30^{\rm o}$ or below 
$-30^{\rm o}$ Galactic latitude, number within $3$~kpc of Earth,
and the percentage of the sky the clumps occupy.}
\label{tab_NFW}
\end{table}

We have created several simulations of a clumpy halo.  In each,
the lower cutoff on the clump mass, $M_{\rm cl,min}$, is a specified
parameter.  Table~\ref{tab_NFW} presents the major results for each
of six halo realizations with $M_{\rm cl,min} = 10^7~M_{\odot}$, for
which 3972 clumps were generated, totaling $\sim$ few 
$10^{11}~M_{\odot}$.  These halo simulations all employ
base case attributes ($\langle \sigma v \rangle_{\chi \bar{\chi}}
= 3 \times 10^{-29}~{\rm cm^3/s}$ and $m_{\chi} = 100$~GeV).

We find that between 250 and 300 clumps will be observable
in this halo, above the CMB anisotropies between 10 and 400~GHz
as described in section 2.3.  None of the NFW clumps is bright
enough to outshine the CMB radiation itself; only the CMB
fluctuations are exceeded.  The NFW clumps in these
realizations occupy slightly less than 0.1\%
of the full sky's solid angle (assuming a minimum pixel
size of $10 \times 10$~arcmin), and  typically several
clumps are within 3~kpc of the Earth.  (Much of this
information in table~\ref{tab_NFW} is intended for comparison
with the corresponding SIS case, given in table~\ref{tab_SIS}
and discussed in section 4.3.)

Of the few hundred observable clumps, between $\sim$ 50
and 75 will reside above $30^{\rm o}$ or
below $-30^{\rm o}$ Galactic latitude, where Galactic
microwave contamination is not problematic.  The locations of all the
clumps in the sky are depicted for halo realization 1 in
figure~\ref{fig_skyNFW}, with crosses indicating the angular size of
the clumps such that half of their total emission comes from inside
that area.  This half-light angle corresponds to a half-light radius,
which we will call $R_{1/2}$, that equals $0.26 r_c$ for an NFW clump.
For clarity, the angular sizes in the figure are
exaggerated by a factor of 10.

\begin{figure}[t]
\vspace*{2.0mm}
\includegraphics[angle=-90,width=8.3cm]{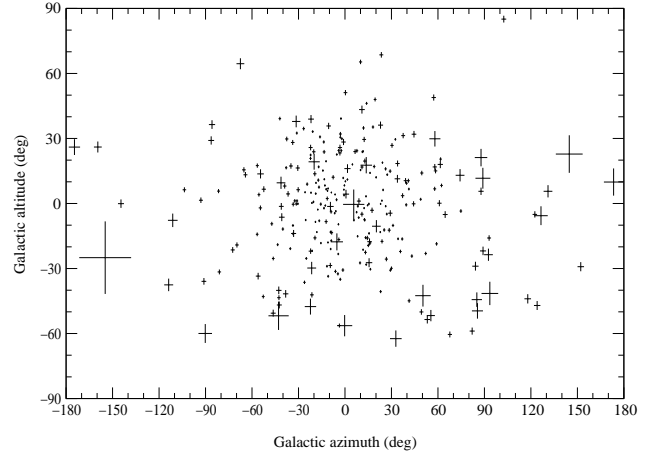}
\caption{A simulated sky of observable NFW dark matter clumps.
Each cross indicates a clump, observable by the criteria laid out
in section 2.3.  The size of the cross indicates the angular size
of the clump such that half its light originates inside the area
shown.  Angular sizes are exaggerated by a factor of 10.}
\label{fig_skyNFW}
\end{figure}

A histogram of the number of clumps
vs. angular size is given in figure~\ref{fig_thnum}.  The figure
shows results for all 6 halo realizations discussed so far, plus
several other cases corresponding to modified parameters
$M_{\rm cl,min}$, $\langle \sigma v \rangle_{\chi \bar{\chi}}$,
and $m_{\chi}$; we modify these in order to cover some of
the possible halos and some of the supersymmetric
parameter space available to the neutralino.
In each case, $\sim$ 100 clumps are
resolvable, unless $m_{\chi}$ is diminished by $\sim$ 100 times
which is too small for any reasonable neutralino model. 
Interestingly, if the clump mass spectrum extends below
$10^7~M_{\odot}$ (in which case there are 23,830 clumps still
totaling
$\sim few \times 10^{11}~M_{\odot}$), or if $\langle \sigma v
\rangle_{\chi \bar{\chi}}$ is significantly
stronger than assumed, many more clumps can be found. A angular
correlation function for this case is studied in \cite{abot02}.

\begin{figure}[t]
\vspace*{2.0mm}
\includegraphics[angle=-90,width=8.3cm]{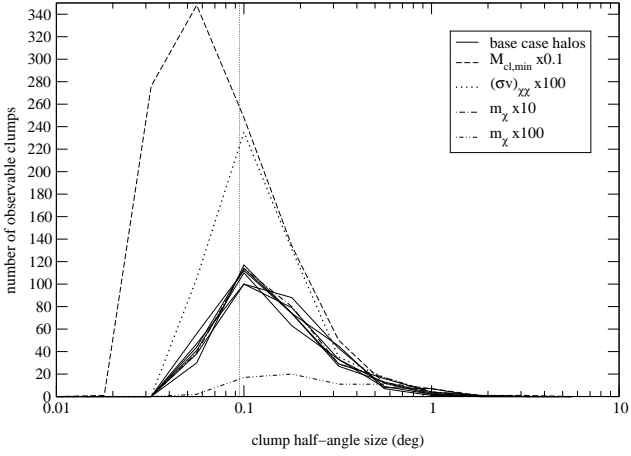}
\caption{Histogram showing the half-light angular size of NFW
clumps in several halo realizations, as used in
figure~\ref{fig_skyNFW}.  Solid lines are for halo realizations
$1 - 6$, where $M_{\rm cl,min} = 10^7~M_{\odot}$.  The dashed
line is for $M_{\rm cl,min} = 10^6~M_{\odot}$; the dotted line
is for $\langle \sigma v \rangle_{\chi \bar{\chi}}$ increased
100 times; the dot-dashed and dot-dot-dashed lines are for
$m_{\chi}$ increased by 10 and 100 times, respectively.
The thin dotted vertical line shows the angular scale of a
$10 \times 10$~arcmin pixel for reference.}
\label{fig_thnum}
\end{figure}

\section{SIS Clumps}

The SIS density profile was given in eq.~\ref{SIS_profile}.
The steep cusp  diverges at the center and
makes it necessary to define $R_{\rm min}$, the radius inside
which neutralino annihilations are so rapid that $n_{\chi} (r \le
R_{\rm min}) = n_{\chi}(R_{\rm min})$ remains constant.
$R_{\rm min}$ is found by setting the cusp-forming timescale
equal to the $\chi \bar{\chi}$
annihilation timescale \citep{bgz92}, so that
\begin{equation}
R_{\rm min} = R_{\rm cl}\, \langle \sigma v
\rangle_{\chi \bar{\chi}}^{1 \over 2} \left[ {n_{\rm halo}
\over 4 \pi G m_{\chi} } \right]^{1 \over 4}\,,
\label{Rmin}
\end{equation}
where $n_{\rm halo}$ is the halo DM density at the location
of the clump.

In the absence of other interactions, one would integrate
$j_{\nu}$ from eq.~\ref{jnu} over the volume of the clump, and find
that the dominant part of the flux comes from the vicinity
of $R_{\rm min}$.  However, due to the high density
at the center of an SIS clump, there are other important
factors to include. We discuss each factor in turn:

\subsection{Relevant Interactions}

(1) {\it Inverse Compton scattering} of electrons
and positrons against the synchrotron photons created by the
electrons and positrons is the most important correction to the
electron-positron distribution in space and energy. To see this,
we write the ICS power as ~\citep{rl}
\begin{equation}
{dE_{{\rm ics}} \over dt} = {4 \over 3} \sigma_T c \beta^2
\gamma^2 U_{\rm ph}\,,
\label{dEdtics}
\end{equation}
where $\sigma_T = 6.65 \times 10^{-25}~{\rm cm^2}$ is the Thomson
cross section (which is appropriate in the limit $h \nu \ll m_e
c^2$), $\beta \simeq 1$, and $U_{\rm ph}$ is the energy density in
the photon field.  

The photon population with the greatest energy density controls the
rate of ICS.  For the CMB,
\begin{equation}
U_{\rm cmb} = a T_{\rm cmb}^4\,,
\label{Ucmb}
\end{equation}
where $a = 7.56 \times 10^{-15}~{\rm erg\,cm^{-3}\,K^{-4}}$
and $T_{\rm cmb} = 2.728~{\rm K}$. At the inner cores of the SIS
clumps, we find that  the synchrotron photon population dominates
the photon energy density such that $U_{\rm ph} \simeq U_{\rm
syn} > U_{\rm cmb} $. The synchrotron energy density at distance
$r$ from the clump center can be calculated as the integral of the
emissivity along all lines of sight, $s(r)$, from $r$ to all other
points in the clump. The
energy and space dependence in the equilibrium density of
electrons and positrons can be factorized  as
$dn_e(E_e,r)/dE_e\propto
\sigma_0(r)
\Phi(E)$, following from a similar factorization in the injection
function
$q(E_e,r)$. This leads to 
\begin{equation}
U_{\rm syn}(r) \propto B^2 m_{\chi}^3 I(r)
\end{equation}
where
\begin{equation}
I(r) = \frac{1}{2}\int_{-1}^{1} d cos\theta \int_0^{s_{\rm max}}
\sigma_0 (r(s)) ds \ , 
\end{equation}
and $s_{\rm max} = {r cos\theta
+((r cos\theta)^2 +R_c^2 -r^2)^{1/2}}$.  After some algebraic
steps we find that
\begin{equation}
\sigma_0(r)\propto r^{-\beta}/I(r),
\end{equation}
where $\beta$ is the slope of the radial dependence in the
injection function ($\beta=4$ for isothermal clumps outside the
minimum radius).
The detailed equation is then solved iteratively in order to 
calculate the equilibrium electron positron function, which
gives
\begin{equation}
{dn_e (E_e, r) \over dE_e }\propto E_e^{-5/2} r^{-5/2}\, ,
\label{dndEsim1}
\end{equation}
for an isothermal clump. Note that in the most general case that
we considered numerically, the equation to solve is an integral
equation since the function $I(r)$ contains the unknown equilibrium
distribution.

This holds wherever electron losses are fastest by ICS against
the synchrotron photon field.  In the case of losses being
dominated by synchrotron radiation and ICS against the CMB instead,
the loss mechanism is independent of $r$, and
$dn_e/E_e\propto E_e^{-5/2} r^{-4}$.  Since the two forms of
electron losses have different power laws in $r$ (and the same
power law in $E_e$), there is a transition radius $R_{\rm ics,syn}$
(which is independent of $E_e$) between the two behaviors.

(2) {\it Pair annihilation} between $e^-$ and $e^+$ particles
would have an important role in determining the flux from a
DM source in the absence of the ICS described above.
But in the presence of ICS losses, pair annihilation
makes little difference to the resulting radio flux.
Pair annihilation is still important, because it
reduces the population of the lower energy pairs, which would
otherwise upscatter photons frequently enough to affect
the synchrotron spectrum. 

We can estimate the $e^{\pm}$ lifetime for the
case where electrons annihilate before they lose a
significant fraction of their original energy
via synchrotron or ICS as
\begin{equation}
\tau_{e^+} \simeq \frac{1}{\langle n_{e^-}
\sigma_{e^{\pm}} v_{e^-} \rangle}
\label{tau_annih}
\end{equation}
with $v_e \simeq c$, and the angle brackets indicating an average
over $E_{e^-}$.  (This is equally valid if we interchange all
the plus and minus signs, but we will keep track of signs in order
to distinguish between particles and antiparticles.)

With eq.~(\ref{dndE}), we have
\begin{equation}
\frac{dn_{e^+}}{dE_{e^+}} = \frac{q_{e^+}}{\langle n_{e^-}
\sigma_{e^{\pm}} v_{e^-} \rangle}.
\label{ne_eq}
\end{equation}
The term in angle brackets is computed from the equations
in~\citet{rs} (also consistent with~\citet{cb}).
 $\langle n_{e^-} \sigma_{e^{\pm}} v_{e^-} \rangle$ can be 
expressed as an energy integral which depends on the $e^{\pm}$
distribution that we are trying to determine.  Again, we
find the solution to this non-linear equation via numerical 
iterations. 

Neglecting a slow varying logarithm in the annihilation cross
section, one finds that $n_{e^-}$ is roughly constant and
$\langle \sigma_{e^{\pm}} v_{e^-} \rangle \propto
E_{e^+}^{-1}$.  That is, a given positron of any energy sees
essentially the same distribution of target electrons, so to a
first approximation, it is sufficient to replace the electron
distribution with a single population at energy $E_{e^-} \simeq
m_e c^2$.  By that rationale,
$\tau_{e^+} \sim E_{e^+}$. At low energies
where $dN_e/dE_e \sim E_e^{-3/2}$, this means that $dn_e/dE_e
\sim E^{-1/2}$ and $n_e \sim E^{1/2}$ (where the subscript $e$
refers to either an electron or a positron).

The distribution in space, rather than energy, is easier to
determine.  In order to satisfy eq.~(\ref{ne_eq}) with the
injection $q_e \propto r^{-4}$ set by the isothermal DM profile,
we must have $dn_e/dE_e \propto r^{-2}$.  Then for a region whose
$e^{\pm}$ losses are fastest by pair annihilation, we have
\begin{equation}
{dn_e \over dE_e} \propto E_e^{-1/2} r^{-2}\,.
\label{dndEsim_annih}
\end{equation}

We define $R_{\rm ann,ics}$ as the transition radius between the
$r^{-2}$ pair annihilation distribution and the $r^{-5/2}$ ICS
distribution.  If $R_{\rm ann,ics} > R_{\rm
min}$, pair annihilation dominates the central region over ICS. 
In our halo and for most choices of $m_{\chi}$,
$R_{\rm ann,ics} < R_{\rm min}$ for electrons radiating
at frequencies of interest. The distribution flattens
at the center before the density can get high enough
to create a pair annihilation dominated region. At lower
frequencies, pair annihilation becomes important and limits the
number of low energy pairs which become unable to effect the
synchrotron spectrum.

(3) {\it Synchrotron self absorption} (SSA) occurs when the 
electrons in the magnetic field can absorb the energy of
the synchrotron photons. The per-unit-length absorption
coefficient for SSA can be written as:
\begin{equation}
\alpha_{\nu} = -{c^2 \over 8 \pi \nu} {dE_{{\rm syn}} \over dt}
\left[ {\partial \over \partial \nu}
\left( {1 \over \nu} {dn_e \over d\nu} \right) \right]\, ,
\label{alpha_ssa}
\end{equation}
where the photon frequency is related to the electron energy via
$\nu_{max}$.
 SSA absorbers  of a particular photon are electrons of
similar energy to the original synchrotron emitter which made the
photon. As such, SSA is only effective below some
$\nu_{\rm crit}$, where the corresponding electrons are more
numerous.

The source function for an absorbing source is
$S_{\nu} = j_{\nu} / \alpha_{\nu}$, and the
optical depth is
\begin{equation}
\tau_{\nu} = \int \alpha_{\nu}\,dz\,,
\label{taudef}
\end{equation}
where $z$ is a line-of-sight coordinate.  The flux
density from such a source at distance $d$ from the Earth
is obtained by summing over each line of sight through the
DM clump:
\begin{equation}
I_{\nu} = {1 \over 4 \pi d^2} \int_0^{R_{cl}} 2 \pi 
b S_{\nu}(1-e^{-\tau_{\nu}})\,db\,\,\,
{{\rm erg} \over {\rm cm^2\,s\,Hz}}\,.
\label{Inu_Snu}
\end{equation}

The SSA cutoff frequency $\nu_{\rm crit}$ depends in part on the absorber
density, and without ICS or $e^{\pm}$ pair annihilation effects,
SSA would be the most important correction to the calculated
flux.  SSA has been included in the calculations performed here,
but since the central absorber population is diminished
by ICS, it is of secondary importance.

\begin{figure}[t]
\vspace*{2.0mm}
\includegraphics[angle=-90,width=8.3cm]{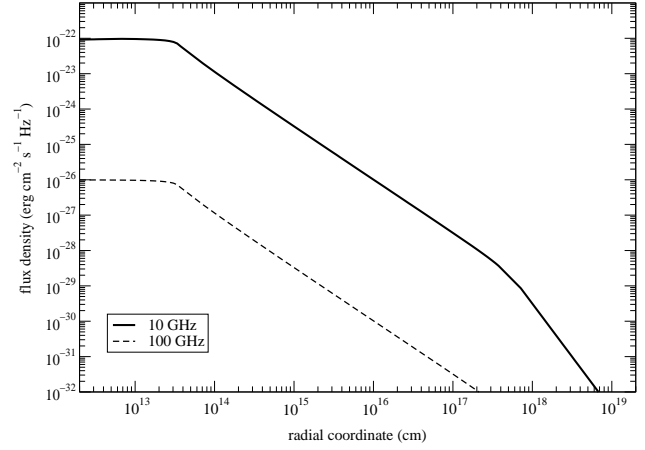}
\caption{Radial flux profile for an example SIS clump,
integrated along the line of sight through the clump.
This plot gives the relative intensities found by pointing
toward different parts of the clump, given sufficient
angular resolution.  The solid line is for 10~GHz and
the dashed line is for 100~GHz.}
\label{fig_fluxr}
\end{figure}

\begin{figure}[t]
\vspace*{2.0mm}
\includegraphics[angle=-90,width=8.3cm]{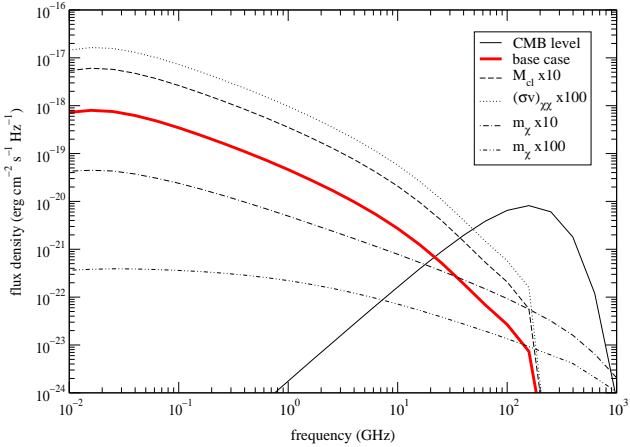}
\caption{Spectrum of synchrotron emission from an example
SIS clump chosen to lie at Galactocentric coordinates
[-4,0,0] kpc.  The thick solid line gives the base case,
$M_{\rm cl} = 10^8~M_{\odot}$,
$\langle \sigma v \rangle_{\chi \bar{\chi}} = 3 \times
10^{-29}~{\rm cm^3/s}$, and $m_{\chi} = 100~{\rm GeV}$.
The dashed line increases $M_{\rm cl}$ by 10 times;
the dotted line increases $\langle \sigma v \rangle_{\chi
\bar{\chi}}$ by 100 times; the dot-dashed line increases
$m_{\chi}$ by 10 times; the dot-dot-dashed line increases
$m_{\chi}$ by 100 times. The thin solid line gives the
intensity of the CMB within the $10 \times 10$~arcmin
pixel, for comparison.}
\label{fig_specSIS}
\end{figure}

\subsection{Results}

The synchrotron spectrum of DM clumps is predominantly altered
by ICS off synchrotron photons. 
Figure~\ref{fig_fluxr} shows an example of a SIS clump
profile that shows the flattened central region.
A characteristic size for the clump, $R_{1/2}$, can be defined by 
the radius inside which half of the synchrotron radiation
originates. 

Figure~\ref{fig_specSIS} shows the spectrum of synchrotron emission
from a SIS clump in the same position as the NFW clump of
figure~\ref{fig_specNFW}.  The thick solid line gives the base
case, $M_{\rm cl} = 10^8~M_{\odot}$,
$\langle \sigma v \rangle_{\chi \bar{\chi}} = 3 \times
10^{-29}~{\rm cm^3/s}$, and $m_{\chi} = 100~{\rm GeV}$.
The dashed line increases $M_{\rm cl}$ by 10;
the dotted line increases $\langle \sigma v \rangle_{\chi
\bar{\chi}}$ by 100 times; the dot-dashed line increases
$m_{\chi}$ by 10 times; the dot-dot-dashed line increases
$m_{\chi}$ by 100 times. The thin solid line gives the
intensity of the CMB within the $10 \times 10$~arcmin
pixel, for comparison.

In order to compare the SIS with the NFW clumps, we used the same
halo simulations as in the NFW case: realizations $1 - 6$ with 3972
clumps ($M_{\rm cl,min} = 10^7~M_{\odot}$), and another
realization with 23,830 clumps ($M_{\rm cl,min} =
10^6~M_{\odot}$), all with $\sim$ few $\times 10^{11}~M_{\odot}$
total. For a typical clump mass of $10^8~M_{\odot}$, a SIS clump
can outshine a NFW clump by about $10^5$ times in the microwave. 

\begin{figure}[t]
\vspace*{2.0mm}
\includegraphics[angle=-90,width=8.3cm]{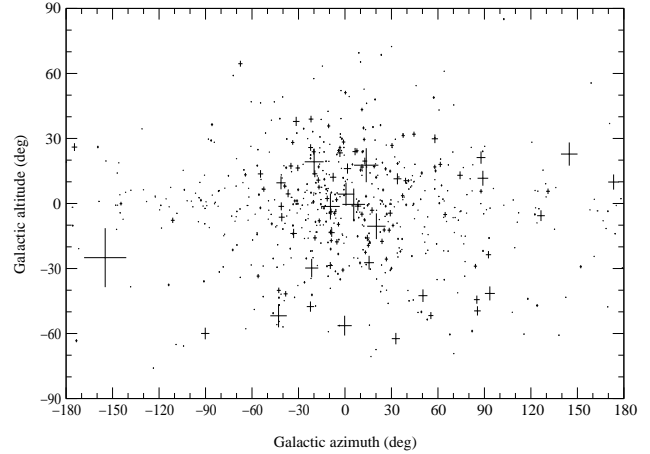}
\caption{A simulated sky of observable SIS dark matter clumps.
Each mark indicates a clump, observable by the criteria laid out
in section 2.3.  The size of the cross indicates the angular size
of the clump such that half its light originates inside the area
shown.  Angular sizes are exaggerated by a factor of 5000.}
\label{fig_skySIS}
\end{figure}

The visible SIS clumps for halo realization 1 are shown in
figure~\ref{fig_skySIS}, here with angular sizes based on
$R_{1/2}$ exaggerated by a factor of 5,000 (none of these
clumps is resolvable without interferometry).
In table~\ref{tab_SIS}, we list some results of the SIS case.
First, we find that SIS clumps (in an NFW halo) tend to produce
$600 - 700$ observable sources, which is a few times more
than the NFW clump case (compare tables~\ref{tab_NFW}
and~\ref{tab_SIS}).  The combined sources occupy only slightly
less solid angle than with NFW, mostly due to the small angular
size of SIS clumps.  But unlike the NFW case, $\sim$20\% of the SIS
clumps not only outshine the CMB anisotropies, but outshine the CMB
altogether. A similar percentage of observable clumps
is located outside the direction of the Galactic disk.  Finally,
the table gives several observable clumps within 3~kpc
of the Earth, which can be resolved  
(as in figure~\ref{fig_fluxr}) by long baseline
interferometers.

\begin{table}
\begin{center}
\begin{tabular}{|l|rrrrrr|} \hline
{\bf SIS} & {\bf 1} & {\bf 2} & {\bf 3} & {\bf 4} 
& {\bf 5} & {\bf 6} \\ \hline
observable & 636 & 632 & 676 & 656 & 682 & 616 \\
outshine CMB & 54 & 34 & 38 & 37 & 37 & 40 \\
$> 30^{\rm o}$~latitude & 130 & 155 & 144 & 129 & 140 & 135 \\
within 3~kpc & 9 & 8 & 2 & 5 & 5 & 7 \\
resolvable & 212 & 184 & 209 & 190 & 210 & 201 \\
resolve $> 30^{\rm o}$ & 48 & 53 & 51 & 49 & 54 & 59 \\
\% of sky & 0.04 & 0.04 & 0.05 & 0.04 & 0.05 & 0.04 \\
\hline
\end{tabular}
\end{center}
\caption{Some results for SIS clumps in six halo realizations, each
set with $M_{\rm cl,min} = 10^7~M_{\odot}$, yielding 3972 clumps.
The following information on {\it observable} clumps is given:
number of them, number which outshine the CMB (rather than just
the CMB anisotropies), number above $30^{\rm o}$ (or below
$-30^{\rm o}$) Galactic latitude, number within $3$~kpc of Earth,
number resolvable by 10~km interferometry such as the VLA,
number resolvable by the VLA above $30^{\rm o}$ Galactic latitude,
and the percentage of the sky the clumps occupy.
\label{tab_SIS}}
\end{table}

Table~\ref{tab_nfwsis} gives a comparison of the NFW and SIS
clump profiles, by reporting the number of observable clumps
in each case, while also varying $m_{\chi}$,
$\langle \sigma v \rangle_{\chi \bar{\chi}}$, or
$M_{\rm cl,min}$.  The results show that changing
these parameters has different outcomes for NFW and SIS.
For example, increasing $m_{\chi}$ by a factor of 100
lowers the spectrum for most relevant $\nu$ bands (see
figure~\ref{fig_specNFW}); therefore, NFW clumps get dimmer and fewer
are seen.  On the other hand, the SIS clumps are bright enough
to avoid that worry, and the increase in $m_{\chi}$ also
increases the maximum electron energy.  This allows strong
sources even in weak magnetic fields via eqs.~\ref{dEdtsyn}
and~\ref{numax}, raising the number of observable clumps.

\begin{table}
\begin{center}
\begin{tabular}{|l|lr|} \hline
{\bf case} & {\bf NFW} & {\bf SIS} \\ \hline
base case & 282 & 636 \\
$\langle \sigma v \rangle_{\chi \bar{\chi}} \times 100$ & 528 & 671 \\
$m_{\chi} \times 10$ & 284 & 1126 \\
$m_{\chi} \times 100$ & 66 & 917 \\
$M_{\rm cl,min} \times 0.1$ & 1078 & 3503 \\
\hline
\end{tabular}
\end{center}
\caption{Comparison of NFW and SIS clumps.  The base case refers
to $m_{\chi} = 100$~GeV, $\langle \sigma v \rangle_{\chi \bar{\chi}}
\simeq 3 \times 10^{-29}~{\rm cm^3/s}$, and
$M_{\rm cl,min} = 10^7~M_{\odot}$.  The other cases modify these
parameters as stated.  The last row of the table uses a different
halo realization than the other rows.
\label{tab_nfwsis}}
\end{table}

As another example,
changing $\langle \sigma v \rangle_{\chi \bar{\chi}}$ has
little effect on SIS clumps, because they have an overdensity
of electrons and positrons emitting anyway.  But for NFW, where
the clumps are typically near the CMB anisotropy level, changing
the $\chi \bar{\chi}$ cross section affects the fluxes
significantly. Both NFW and SIS cases see similar growth
in the number of observable clumps when the clump mass spectrum is
extended downward.

\section{Conclusion}

In this paper, we studied the synchrotron signature of neutralino
annihilation in the presence of the Galactic magnetic field.
If hierarchical CDM clustering  extends down to mass scales
smaller than dwarf galaxies, these DM  clumps are detectable in
microwaves over the fluctuations in the CMB.

If the clumps have nearly NFW profiles like the Galaxy, then
the clumps have finite angular sizes.  NFW clumps have 
synchrotron spectrum from non-thermal electrons with negligible
ICS and self-absortion.  NFW clumps will be large enough and
numerous enough that several should be resolvable in angular
profile by CMB anisotropy experiments.  The light profile
will follow $r^{-1}$ inside a core radius. The
statistics of the clumps distribution should be consistency with
figure~\ref{fig_thnum} and table~\ref{tab_NFW}.

SIS clumps are brighter and smaller in angular size.
 The clumps have a distinctive spectrum due
to a combination of ICS and pair annihilation features, in
a particular radial structure.  The spectrum can be checked at
lower frequencies also, because these clumps are very bright and
may outshine other radio backgrounds.  A few clumps
should be close enough to display a distinctive radial profile by
interferometric means.

In either case, one should make use of CMB anisotropy studies to
search for neutralino annihilation in DM clumps.  
Many clumps should have signal above $2
\times 10^{-5}$ times the 2.728~K Planck function.  The sources
are usually small and occupy a small fraction of the sky, so that
a full sky high resolution microwave survey maybe necessary.  The
upcoming MAP and Planck missions will provide that.  Even in
smaller data sets such as the BOOMERANG, MAXIMA, and DASI,  
it is possible that  clumps can be detected.

Upon finding synchrotron  source clumps or hot pixels, one final
check can prove their DM nature.  The gamma ray and neutrino
emission from them must fit the expected levels, and must be
independent of position in the Galaxy, unlike the synchrotron
which follows the Galaxy's magnetic field.  This combination of
signatures makes the sources unique, and aids in extracting
specific information about them:  $M_{\rm cl}$, $m_{\chi}$, and
$\langle \sigma v \rangle_{\chi \bar{\chi}}$.

One interesting point to note is that for reasonable choices of
the appropriate parameters, we find that several hundred or more
clumps are typically observable in either profile.  If new CMB
experiments fail to detect them, then we have learned that either
neutralinos are not the dark matter, or CDM halo
clumps do not exist.  However, if they are found, they offer
a valuable opportunity to learn more about CDM clustering, neutralino
properties like mass and cross section, and possibly the Galactic
magnetic field structure.

\begin{acknowledgements}

This work was supported in part by the NSF through grant
AST-0071235 and DOE grant DE-FG0291-ER40606 at the University of
Chicago and at the Center for Cosmological Physics by grant NSF
PHY-0114422.

\end{acknowledgements}


\begin{thebibliography}{99}

\bibitem[Aloisio et al.(2002a)]{AloGC} Aloisio, R., Blasi, P.,
Olinto, A.V.,  in preparation.


\bibitem[Aloisio et al.(2002b)]{abot02} Aloisio, R., Blasi, P.,
Olinto, A.V., Tasitsiomi, A., in preparation.


\bibitem[Berezinsky, Gurevich, and Zybin(1992)]{bgz92} Berezinsky, V.,
Gurevich, A.V.,  and  Zybin, K.P., Phys. Lett.  B 294, 221, 1992.

\bibitem[Berezinsky, Bottino, and Mignola(1994)]{bbm94} Berezinsky, V.,
Bottino, A.,  and  Mignola, G. Phys. Lett.  B 325, 136, 1994.

\bibitem[Bergstr\"om et al.(1999)]{begu99} Bergstr\"om, L., Edsj\"o, J.,
Gondolo, P.,  Ullio,  P.,  Phys. Rev. D  59, 043506, 1999.

\bibitem[de Bernardis et al.(2000)]{boom00} de Bernardis, P. et al.
(BOOMERANG collaboration), Nature 404, 955, 2000; de Bernardis, P.
et al., astro-ph/0011469.

\bibitem[Blasi and Sheth(2000)]{bs00}
Blasi, P., and Sheth, R. K.,   Phys. Lett.  B  486, 233, 2000.

\bibitem[Calc\'{a}neo--Rold\'{a}n and Moore(2001)]{cm01}
Calc\'{a}neo--Rold\'{a}n, C. and Moore, B., astro-ph/0010056.

\bibitem[Chiba(2001)]{mc01} Chiba, M., astro-ph/0109499.

\bibitem[Coppi and Blandford(1990)]{cb} Coppi, P.S. and Blandford,
R.D., MNRAS 245, 453, 1990.

\bibitem[Fixsen et al.(1996)]{fix96} Fixsen, D.J., Cheng, E.S.,
Gales, J.M., Mather, J.C., Shafer, R.A., Wright, E.L., Ap.\ J.
473, 576, 1996.

\bibitem[Fixsen et al.(1997)]{fix97} Fixsen, D.J., Hinshaw, G.,
Bennett, C.L., Mather, J.C., Ap.\ J. 486, 623, 1997.

\bibitem[Fomalont et al.(1993)]{vla93} Fomalont, E.B., Partridge, R.B.,
Lowenthal, J.D., Windhorst, R.A., Ap.\ J. 404, 8, 1993.

\bibitem[Ghigna et al.(1998)]{gmglqs}
Ghigna, S., Moore, B., Governato, F., Lake, G., Quinn, T., Stadel, J.,
   MNRAS   300, 146, 1998.

\bibitem[Gondolo(2000)]{g00}  Gondolo, P., preprint hep-ph/0002226.

\bibitem[Gondolo and Silk(1999)]{gs99} Gondolo, P.  and  Silk, J., Phys.\
Rev.\ Lett.\   83, 1719, 1999.

\bibitem[Hanany et al.(2000)]{max00} Hanany, S. et al., Ap. \ J.
545, L5, 2000.

\bibitem[Hill(1983)]{hill} Hill, C.T., Nucl. Phys. B224, 469
(1983).

\bibitem[Hinshaw(2000)]{map00} Hinshaw, G., astro-ph/0011555.

\bibitem[Jungman, Kamionkowski and Griest(1996)]{jkg96}
Jungman, G.,  Kamionkowski, M., and Griest,  K., Phys.\ Rep.\   267, 195,
1996.

\bibitem[Navarro, Frenk and  White(1996, 1997)]{nfw}  Navarro, J.F.,
Frenk, C.S.,  and White,  S.D.M., Ap.\ J.  462,
   563, 1996; ibid.\  490, 493, 1997.

\bibitem[Partridge et al.(1997)]{vla97} Partridge, R.B., Richards, E.A.,
Fomalont, E.B., Kellerman, K.I., Windhorst, R.A., Ap.\ J. 483, 38, 1997.

\bibitem[Rybicki and Lightman(1979)]{rl} Rybicki, G.B. and Lightman,
A.P., ``Radiative Processes in Astrophysics'', 1979.

\bibitem[Salucci(2000)]{sal00} Salucci, (2000).

\bibitem[Smoot et al.(1992)]{smoot92} Smoot, G.F. et al.
(COBE DMR collaboration), Ap.\ J. 396, L1, 1992.

\bibitem[Stanev(1997)]{st97}  Stanev, T. , Astrophys.
J. 479,  290,  1997.

\bibitem[Svennson(1982)]{rs} Svensson, R., Ap.J. 258, 321, 1982.

\bibitem[Tegmark et al.(2000)]{teho00} Tegmark, M.,
Eisenstein, D.J., Hu, W., and de Oliveira-Costa,  A., Ap.J. 530, 133,
2000.

\bibitem[Tyler(2002)]{ct} Tyler, C., in preparation.

\bibitem[de Zotti et al.(1999)]{planck} de Zotti, G. et al.,
astro-ph/9902103.

\end{thebibliography}
\end{document}